# Surface plasmon resonance spectroscopy of single bowtie nano-antennas using a differential reflectivity method

M. Kaniber[1,*], K. Schraml[1], A. Regler[1], J. Bartl[1], G. Glashagen[1], F. Flassig[1], J. Wierzbowski[1] & J. J. Finley[1,2]

[1] Walter Schottky Institut and Physik Department, Technische Universität München, Am Coulombwall 4, 85748 Garching b. München, Germany

[2] Nanosystems Initiative Munich (NIM), Schellingstraße 4, 80799 München, Germany.

* Correspondence and requests for materials should be addressed to M.K. (kaniber@wsi.tum.de)

**We report on the structural and optical properties of individual bowtie nanoantennas both on glass and semiconducting GaAs substrates. The antennas on glass (GaAs) are shown to be of excellent quality and high uniformity reflected by narrow size distributions with standard deviations for the triangle and gap size of $\sigma_s^{glass} = 4.5\,\text{nm}$ ($\sigma_s^{GaAs} = 2.6\,\text{nm}$) and $\sigma_g^{glass} = 5.4\,\text{nm}$ ($\sigma_g^{GaAs} = 3.8\,\text{nm}$), respectively. The corresponding optical properties of individual nanoantennas studied by differential reflection spectroscopy show a strong reduction of the localised surface plasmon polariton resonance linewidth from $0.21\,\text{eV}$ to $0.07\,\text{eV}$ upon reducing the antenna size from $150\,\text{nm}$ to $100\,\text{nm}$. This is attributed to the absence of inhomogeneous broadening as compared to optical measurements on nanoantenna ensembles. The inter-particle coupling of an individual bowtie nanoantenna, which gives rise to strongly localised and enhanced electromagnetic hotspots, is demonstrated using polarization-resolved spectroscopy, yielding a large degree of linear polarization of $\rho_{max} \sim 80\%$. The combination of highly reproducible nanofabrication and fast, non-destructive and non-contaminating optical spectroscopy paves the route towards future semiconductor-based nano-plasmonic circuits, consisting of multiple photonic and plasmonic entities.**

## Introduction

Single metal nanoparticles [1], nanoparticle dimers [2] or even nanoparticle arrays [3] are well known to concentrate visible [4], infrared [5] and microwave [6] radiation from the far-field into sub-wavelength sized optical volumes whilst simultaneously giving rise to strong electric field enhancements on the order of $10^3 - 10^4$ [7] [8]. In particular optical antennas [9] such as bowtie nanoantennas have been shown to provide besides extraordinarily high field enhancements [10], also directionality [11], broadband spectral responses [12], local electrical control [13] with potential for tunability [14], highly efficient electro-optical driving [15] and full polarization control [16]. Amongst others, such systems found already applications in surface enhanced Raman spectroscopy [17], ultra-high resolution lithography [18] and microscopy [19], bio-chemical sensing [20] [21], spontaneous emission control [22] and enhancement [23] [24], non-linear optics [25] [26] and solar energy conversion [27].

Chemical synthesis [28] of plasmonic nanostructures is well established and widely-used since sophisticated and expensive equipment is not required to produce large amounts of plasmonic nanoparticles. However, nano-lithography techniques offer much higher flexibility in controlling and deterministically designing the optical properties of plasmonic nanostructures. For example, it is possible to tailor the localised surface plasmon polariton resonance via precise adjustment of size [29] [30] and shape [31], as well as the polarization of the scattered photons via the antenna geometry [16]. Moreover, the exact control of the particle location and density during the lithography process enables to switch on radiative coupling in arrays of nanoparticles [32] and, thus, give rise to multipolar surface plasmon modes [33] and collective surface lattice resonances [34]. This proves crucial to design novel properties such as magnetic polarizability [35], negative-refractive indices [36] or phase-gradients [37] in metasurfaces [38].

Many spectroscopy techniques for studying single plasmonic nanostructures have been established in recent years [39]. Examples include scanning near-field optical microscopy [40], attenuated total internal reflection [4], extinction or transmission experiments [2] and dark field spectroscopy [41]. However, the majority of those methods either demand expensive equipment, require specially designed samples or contaminate their surface. Therefore, a reliable, fast, non-destructive and cheap measurements method with high spatial resolution would be highly attractive for determining the optical properties of the individual plasmonic nanostructures on a future semiconductor-based plasmonic nano-circuit [42] [43].

Here, we present a systematic and comprehensive study of the structural and optical properties of individual, lithographically defined bowtie nanoantennas [12] on glass and semiconducting GaAs substrates using differential reflection spectroscopy. Therefore, we fabricated antennas with sizes $100\text{nm} \leq s_0 \leq 150\text{nm}$, feed-gaps down to $5\text{nm}$ and tip radii $r_c = 14 \pm 5\text{nm}$ using electron beam lithography [30]. Scanning electron microscopy yields narrow distributions of triangle size and gap size on glass (GaAs) substrates with standard deviations of $\sigma_s^{\text{glass}} = 4.5\text{nm}$ ($\sigma_s^{GaAs} = 2.6nm$) and $\sigma_g^{\text{glass}} = 5.4\text{nm}$ ($\sigma_g^{GaAs} = 3.8nm$), respectively, indicating highly uniform and reproducible nanofabrication. The corresponding optical properties of individual bowtie nanoantennas are investigated using high-spatial resolution, differential reflection spectroscopy, demonstrating the linear (cubic) dependence of the surface plasmon resonance energy $E_{res}$ on the triangles size (gap size) [12] [30]. Comparison between measurements on single and ensembles of bowtie nanoantennas [30] show clear indications of inhomogeneous broadening [4], varying between $0.07\,\text{eV}$ and $0.21\text{eV}$ for $s_0 = 100\text{nm}$ and $s_0 = 150\text{nm}$, respectively. Finally, we study the inter-particle coupling between the two nano-triangles forming the bowtie nanoantenna using polarization-resolved spectroscopy. Those measurements show strongly linearly polarized emission along the main axis of the antenna for the coupled mode with a degree of polarization up to $\rho_{max} \sim 80\%$. Our results are contrasted with studies on semiconductor GaAs substrates and all experiments are shown to be in excellent agreement with numerical simulations [44].

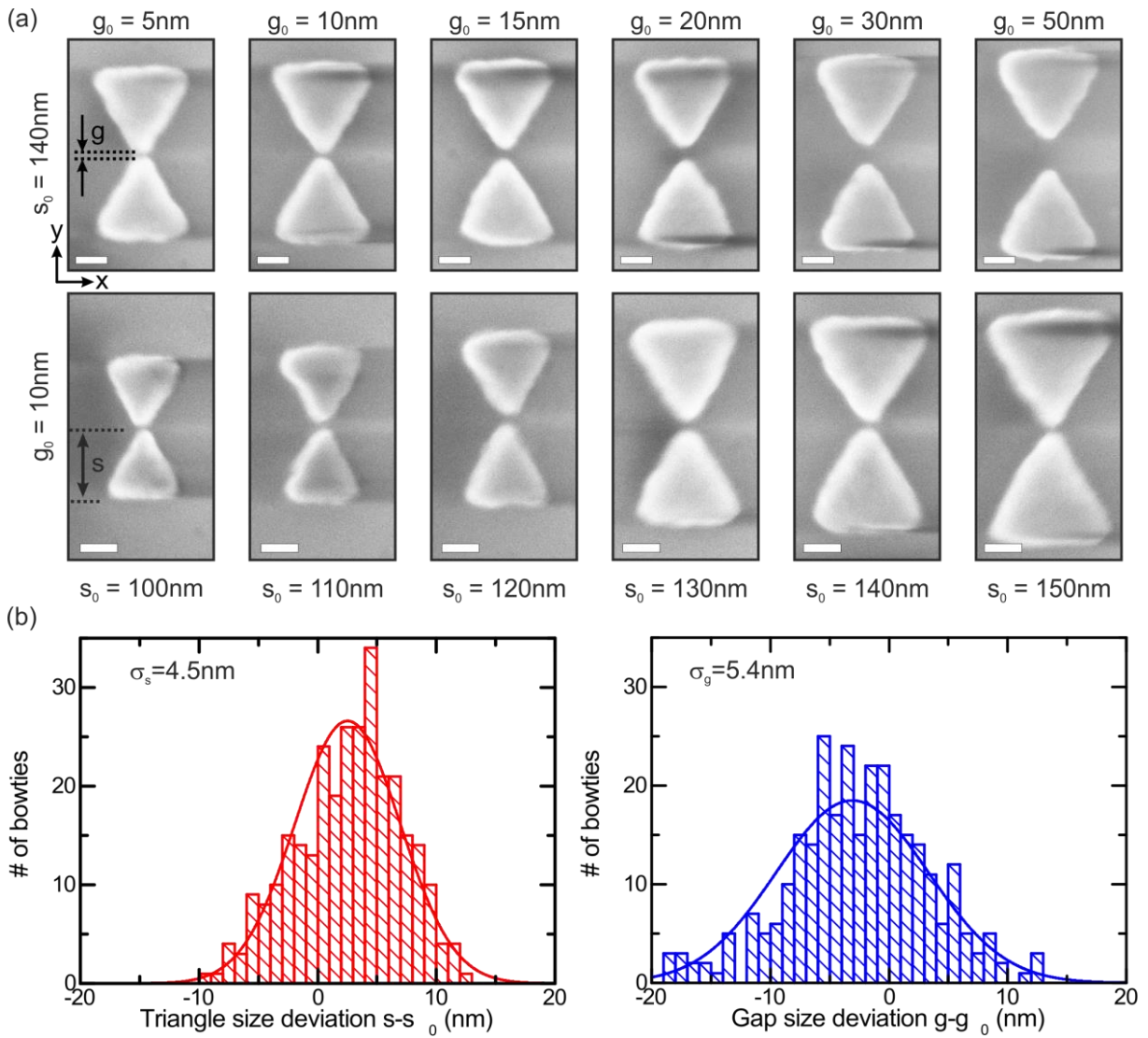

**Figure 1 (a)** Top row: Scanning electron microscopy images of individual bowtie nanoantennas on a glass substrate for $s_0 = 140nm$ as a function of nominal gap size $5nm < g_0 < 50nm$ from left to right, respectively. Bottom row: Scanning electron microscopy images of individual bowtie nanoantennas on a glass substrate for $g_0 = 10nm$ as a function of nominal triangle size $100nm < s_0 < 150nm$ from left to right, respectively. Scale bar, $50nm$. **(b)** Left panel: Statistical analysis of the number of bowtie nanoantennas as a function of the triangle size deviation $s - s_0$. Right panel: Statistical analysis of the number of bowtie nanoantennas as a function of the gap size deviation $g - g_0$.

## Results

In Figure 1 (a), we present a selection of scanning electron microscopy images of lithographically defined Au bowtie nanoantennas on a non-conducting glass substrate. The top (bottom) row shows bowtie nanoantennas for constant nominal triangle size $s_0 = 140 nm$ (gap size $g_0 = 10 nm$) and increasing $g_0$ ($s_0$) between $5 nm$ ($100 nm$) and $50 nm$ ($150 nm$) from left to right, respectively. The antenna thickness was kept constant at $t = 35 nm$. The triangles forming the bowtie nanoantenna are of high quality, indicated by their smooth edges and surfaces without observable distortions or fraying. The typical tip radii was found to be $r_c = 14 \pm 5\ nm$. The highly reproducible fabrication process is further supported by the histograms plotted in Figure 1 (b), representing the number of individual bowtie nanoantennas as a function of triangle size deviation $\Delta_s \equiv s - s_0$ and gap size deviation $\Delta_g \equiv g - g_0$ in the left and right panel, respectively. Here, $s$ and $g$ denote the experimentally determined triangle and gap size, respectively, as defined in the leftmost images in Figure 1 (a). We extracted $s$ and $g$ from high resolution scanning electron microscopy measurements for $\sim 300$ nominally identical bowtie nanoantennas without any pre-selection, with $s_0$ and $g_0$ spanning the range given in Figure 1 (a). Both histograms for $\Delta_s$ and $\Delta_g$ are well described by a Gaussian distribution $y(x) = \frac{A}{\sigma\sqrt{2\pi}} exp(-\frac{(x-\mu)^2}{\sigma^2})$ where $\mu$ and $\sigma$ denote the expectation value and the standard deviation, respectively. From the fits of the triangle size and gap size histograms, we obtain narrow distributions indicated by the small values of the corresponding $\sigma_s^{glass} = 4.5 nm$ and $\sigma_g^{glass} = 5.4 nm$. This means in particular that $\sim 95.4\%$ of the fabricated triangles exhibit deviations in triangles size and gap size of less than $2\sigma_s^{glass} = 9 nm$ and $2\sigma_g^{glass} = 10.8 nm$ from the nominal values, respectively. We further note that the shift of both triangle and gap size distributions from $\Delta_{s,g} = 0$, reflected by $\mu_s^{glass} = 3.1 nm$ and $\mu_g^{glass} = -2.4 nm$, can easily be compensated by fine-adjusting the dose during the electron beam lithography. Similar structural investigations for Au bowtie nanoantennas on high-refractive index ($n_{GaAs} = 3.54$ at $T = 297\ K$ and $E_{photon} = 1.3 eV$ [45]), semiconducting GaAs substrates showing even narrower distributions with $\sigma_s^{GaAs} = 2.6 nm$ and $\sigma_g^{GaAs} = 3.8 nm$ are presented in the Supplementary Material, figure SM1. We conclude that we established a highly reproducible lithography process for bowtie nanoantennas with a fabrication accuracy of $\sim 10 nm$, limited by the electron beam lithography system, giving rise to reproducibly fabricated nanoantennas with feature sizes down to $10 nm$.

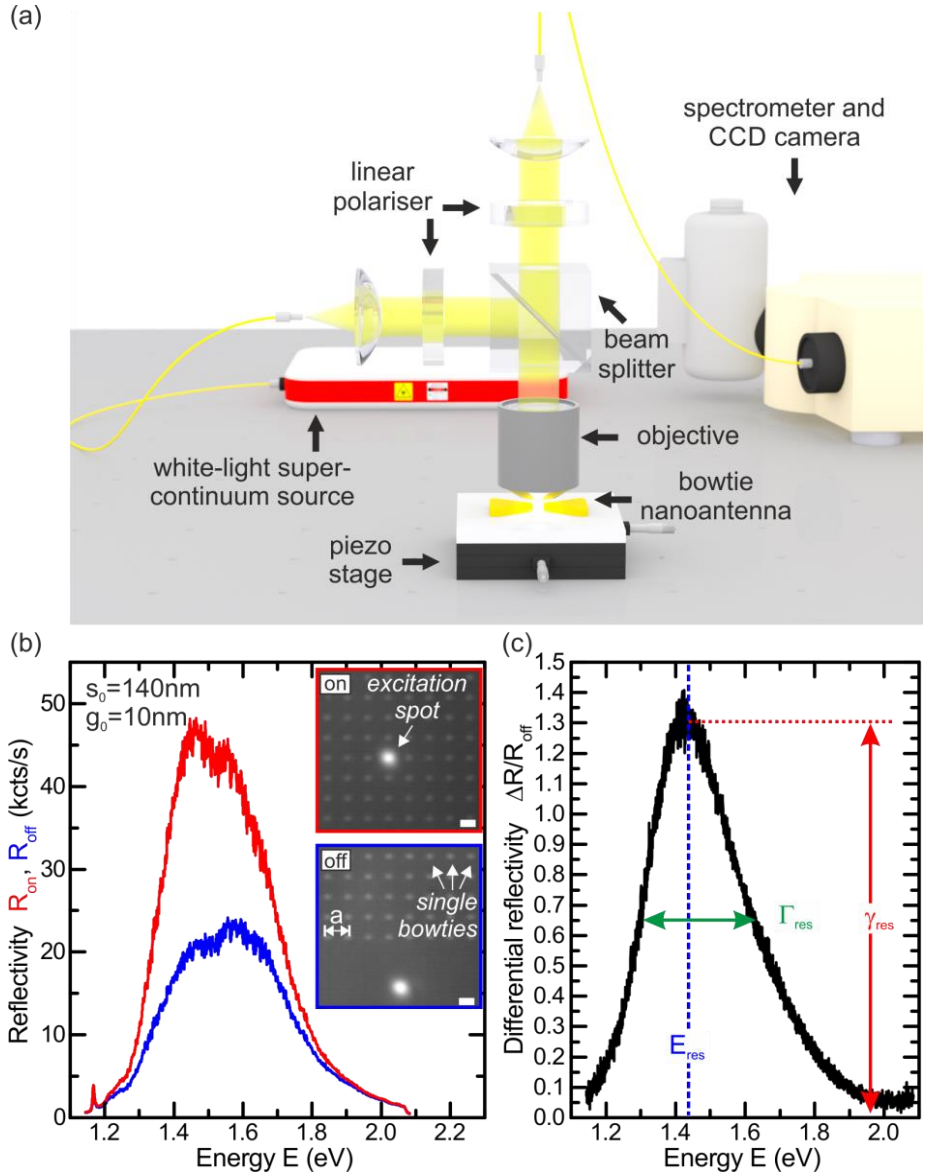

**Figure 2 (a)** Schematic illustration of the differential reflectivity setup. **(b)** Measured reflectivities $R_{on}$ and $R_{off}$ using a diffraction limited white-light super-continuum source spatially positioned on and off a single bowtie nanoantenna with $s_0 = 140 nm$ and $g_0 = 10 nm$ in red and blue, respectively. Insets show light microscopy images of the bowtie array and the white light laser spot. Scale bars, $1 \mu m$. **(c)** Differential reflectivity $\Delta R/R_{off}$ as a function of energy $E$ calculated from the reflectivity spectra shown in **(b)**.

To study the optical response of individual bowtie nanoantennas, we used a home-built confocal microscope that facilitates measurements of the broadband ($\Delta \lambda \sim 400 - 1600 \, nm$) reflectivity of a diffraction limited laser spot generated by a white-light super-continuum source as schematically shown in Figure 2 (a). The excitation beam is reflected from a beamsplitter and focused onto the sample via a microscope objective. The reflected light is collected via the same objective, transmitted through the beamsplitter and guided via an optical fibre to a spectrometer. For more details on the setup and the used optical components we refer to the Methods Section. In order to determine the localised surface plasmon polariton resonance of an individual nanoantenna, we performed two subsequent measurements; first, we measured the reflectivity $R_{on}$ from an individual bowtie nanoantenna as a function of energy $E$ as shown by the red curve in Figure 2 (b). Here, the upper inset depicts a light microscopy image recorded in our setup, which displays the bowtie nanoantennas ($s_0 = 140nm$, $g_0 = 10nm$.) arranged in a periodic array with a lattice constant of $a = 1.5\mu m$ and the white light excitation spot focused on one single antenna. In a second step, we recorded a similar reflectivity spectrum $R_{off}(E)$ from a location spatially displaced from the bowtie nanoantenna array as shown by the lower inset in Figure 2 (b) for reference. The corresponding spectrum $R_{off}(E)$ is plotted in blue. From the measurements of $R_{on}$ and $R_{off}$ we calculate the differential reflectivity $\Delta R/R_{off} \equiv (R_{on} - R_{off})/R_{off}$, which represents a measure for the scattered light

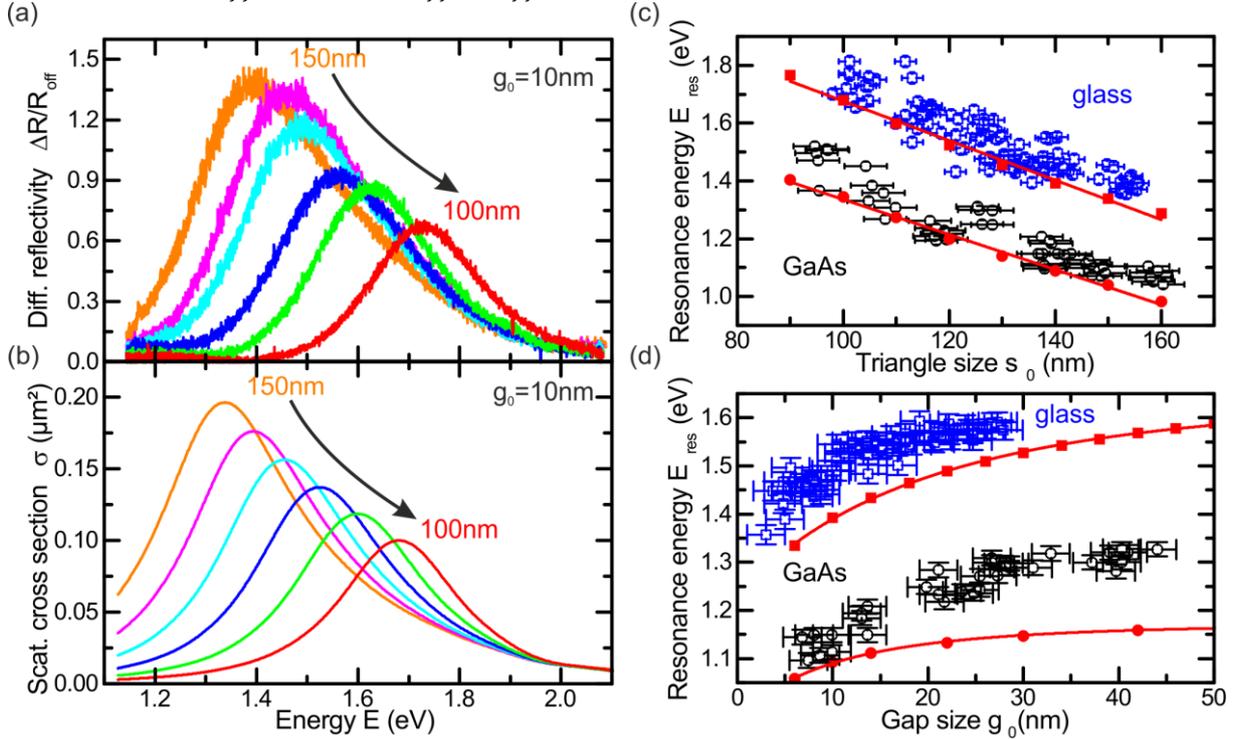

**Figure 3 (a)** Differential reflectivity $\Delta R/R_{off}$ and **(b)** numerically simulated scattering cross section $\sigma$ as a function of energy $E$ for triangle sizes $100nm < s_0 < 150nm$ and $g_0 = 10nm$. **(c)** Localised surface plasmon polariton resonance energy $E_{res}$ as a function of triangle size $s_0$ for $g_0 = 10nm$ on a glass and a GaAs substrate in blue and black, respectively. **(d)** Localised surface plasmon polariton resonance energy $E_{res}$ as a function of gap size $g_0$ for $s_0 = 140nm$ on a glass and a GaAs substrate in blue and black, respectively. Red symbols and curves in **(c)** and **(d)** represent simulation results.

from the bowtie nanoantenna [46]. The $\Delta R/R_{off}$ –spectra determined from the reflectivity measurements shown in Figure 2 (b) is presented in panel (c). We observe a peak-like response with a resonance maximum $\gamma_{res}$ at the resonance energy $E_{res}$, interpreted as the dipolar localised surface plasmon polariton resonance of the investigated bowtie nanoantenna [39]. Furthermore, we can extract from the differential reflectivity spectrum the full width at half maximum $\Gamma_{res}$ and, thus, gain insights into the related plasmon lifetime $T_{pl}$ via $T_{pl} = 2\hbar/\Gamma_{res}$ [41], where $\hbar$ denotes the reduced Planck constant.

In the following, we use our method to systematically study the optical properties of individual bowtie nanoantennas fabricated on both glass and GaAs substrates as a function of $s_0$ and $g_0$. The experimentally obtained $\Delta R/R_{off}$-spectra for $g_0 = 10nm$ and triangle sizes $100nm < s_0 < 150\ nm$ in steps of $\Delta s_0 = 10nm$ are presented in Figure 3 (a) for bowtie nanoantennas on glass. We observe a systematic shift of the localised surface plasmon polariton resonance from $E_{res} = 1.39\ eV$ to higher energies $E_{res} = 1.73eV$ with decreasing $s_0$, attributed to reduced retardation effects of the exciting electromagnetic field and the depolarization field inside the metal particles [47]. The blue-shift in $E_{res}$ is accompanied by a decreasing resonance maximum $\gamma_{res}$ from $\gamma_{res} = 1.37$ to $\gamma_{res} = 0.67$, which is due to a reduction of the geometrical scattering cross-section of the antennas with decreasing $s_0$. In Figure 3 (b), we present corresponding finite-difference time-domain simulations [44] of the scattering cross-section $\sigma$ for bowtie nanoantennas on a glass substrate with $g_0 = 10nm$, $r_c = 20nm$ and varying triangle size $100nm < s_0 < 150nm$. We find increasing $E_{res}$ and decreasing $\gamma_{res}$ with decreasing $s_0$, both in excellent qualitative and quantitative agreement with our experimental results. We compare the measured and simulated data for $E_{res}$ as a function of $s_0$ and $g_0$ in Figure 3 (c) and (d), respectively. Blue (black) symbols denote the experimental results for bowtie nanoantennas on a glass (GaAs) substrate, whilst the red symbols represent the simulation results. In general, we observe a comparable linear (cubic) trend for the $s_0$- ($g_0$-) dependence of bowtie nanoantennas on glass and GaAs with shift-rates for the $s_0$-dependence of $-(6.8 \pm 0.3)meV/nm$ and $-(6.3 \pm 0.2)meV/nm$, respectively. The global red-shift of the GaAs data of $\Delta E \sim 0.3eV$ is due to the increase in refractive index of $\Delta n \sim 2.0$ as compared to glass [30]. The cubic ($\propto g_0^{-3}$) behaviour observed in the gap size dependence in Figure 3 (d) is due to near-field interaction, describing the coupling of the surface plasmons in the two adjacent triangles by a coupling of effective point dipoles [48]. Additional spectra and the corresponding simulated scattering cross-sections for the $g_0$-dependence on glass and the $s_0$- and $g_0$-dependence on GaAs, respectively, are presented in figure SM2. As a consequence, we experimentally studied localised surface plasmon polariton resonances for individual bowtie nanoantennas using differential reflection spectroscopy and obtained excellent agreement with numerical simulations of the scattering cross-sections. This combined experimental-simulation approach enables us to reproducibly design and deterministically control the localised surface plasmon polariton resonance of individual nanoantennas.

As demonstrated in the previous section, the localised surface plasmon polariton resonances of bowtie nanoantennas depend strongly on the triangle size $s_0$ and gap size $g_0$. Even though our fabrication process was shown to be highly reproducible, slight variations in $s_0$ and/or $g_0$ will still result in non-negligible variations of the localised surface plasmon polariton resonances. Therefore, measurements on ensembles of bowtie nanoantennas as investigates in Ref. [30] are expected to show enlarged resonance linewidths $\Gamma_{res}$ due to size- and shape-induced inhomogeneous broadening *[46]*. To test this hypothesis, we compare in Figure 4 (a) two typical differential reflectivity spectra recorded from an individual ($N = 1$) and an ensemble

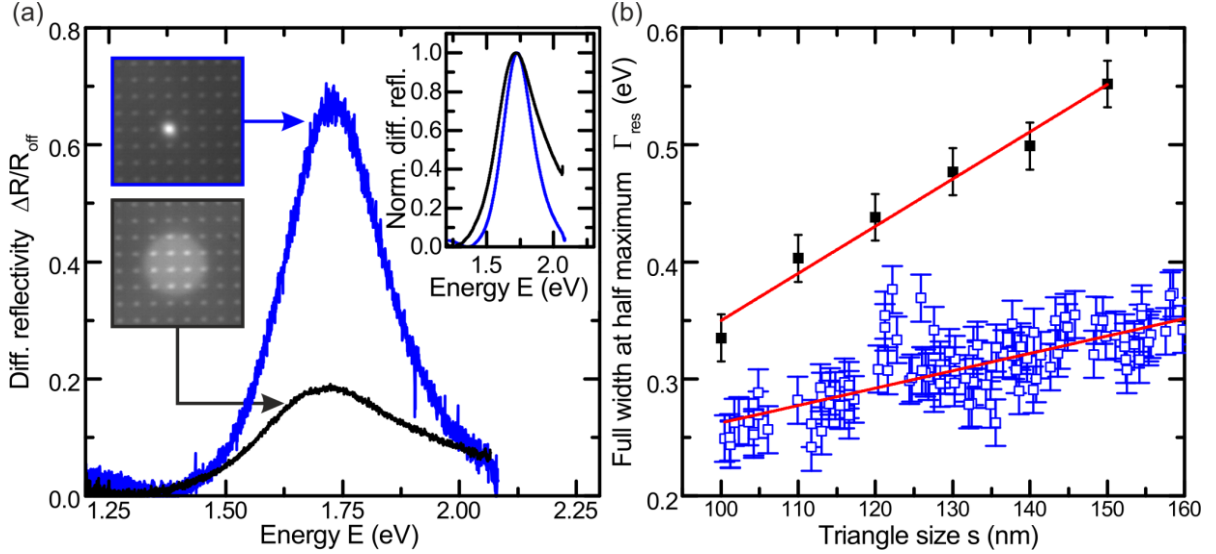

**Figure 4 (a)** Differential reflectivity $\Delta R/R_{off}$ as a function of energy $E$ and gap sizes $g = 10 \pm 3 nm$ for a bowtie ensemble and a single bowtie nanoantenna on a glass substrate plotted in black and blue, respectively. Insets: (Left) Light microscopy images of the bowtie field under illumination with a halogen lamb and the white light super-continuum source in black and blue, respectively. (Right) Same data shown on a normalized differential reflectivity scale. **(b)** Full width at half maximum $\Gamma_{res}$ as a function of triangle size $s$ for bowtie ensembles and single bowtie nanoantennas in black and blue, respectively. Red curves represent linear fits to the data.

($N \sim 12$) of bowtie nanoantennas on a glass substrate in blue and black, respectively. Here, $N$ denotes the number of bowtie nanoantennas excited simultaneously in the differential reflectivity measurements. The left upper and lower insets in Figure 4(a) show white light microscopy images of the bowtie array with the excitation spot of the white light super-continuum source and a halogen lamp for single and ensemble antenna spectroscopy, respectively. The differential reflectivity spectra $\Delta R/R_{off}$ for single and ensemble antenna measurements exhibit a maximum at comparable $E_{res}$, attributed to the localised surface plasmon resonance. However, the corresponding resonance linewidth $\Gamma_{res}$ for the measurement of an individual bowtie nanoantenna is found to be considerably narrower as compared to bowtie nanoantenna ensembles, clearly visible on a normalized differential reflectivity scale as shown in the inset of Figure 4 (a). The larger linewidth for the ensemble measurement is attributed to inhomogeneous broadening.

We systematically investigated this effect by determining $\Gamma_{res}^{N=1}$ of individual bowtie nanoantennas with constant $g = 10 \pm 3 nm$ as a function of measured triangle size $s$. The results of those measurements are plotted the results as blue symbols in Figure 4 (b). The red line represents a linear fit to the data, indicating a systematic broadening of $\Gamma_{res}^{N=1}$ for increasing $s$ from $\Gamma_{res}^{N=1} = 0.26\ eV$ at $s \sim 100nm$ to $\Gamma_{res}^{N=1} = 0.34\ eV$ at $s \sim 150nm$. This observed increase in $\Gamma_{res}^{N=1}$ is attributed to enhanced radiation damping for increasing antenna sizes [49]. Furthermore, we present for comparison differential reflectivity measurements conducted on bowtie nanoantenna ensembles [30] for nominal sizes $100nm < s_0 < 150nm$ with $\Delta s_0 = 10nm$ as black symbols in Figure 4 (b). In addition to the linear increase in $\Gamma_{res}^{N\sim 12}$ with increasing $s_0$ due to enhanced radiation damping, we observe a global offset $\Delta\Gamma_{res} = \Gamma_{res}^{N\sim 12} - \Gamma_{res}^{N=1}$ for the ensemble measurements attributed to inhomogeneous broadening, which varies between $\Delta\Gamma_{res} = 0.07eV$ and $\Delta\Gamma_{res} = 0.21eV$ for $s \sim 100nm$ and $s \sim 150nm$, respectively. Altogether our results demonstrate the impact of small variations in triangle size $s$ and gap size $g$ on the localised surface plasmon polariton resonance of bowtie nanoantennas despite the high fabrication accuracy achievable with state-of-the-art nanotechnology.

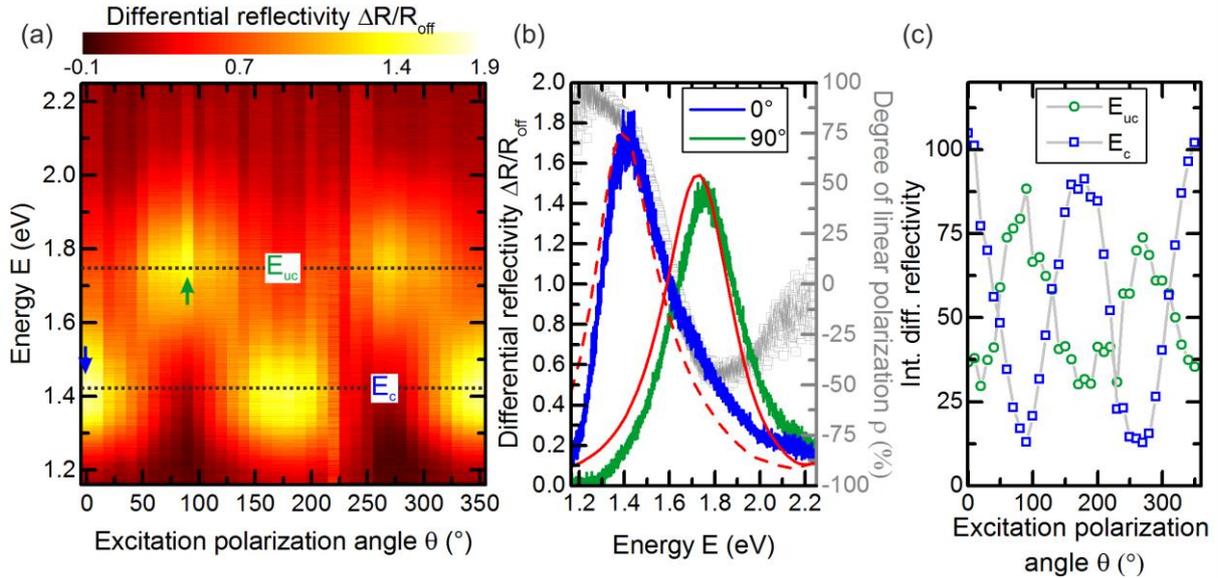

**Figure 5 (a)** Differential reflectivity $\Delta R/R_{off}$ of a single bowtie nanoantenna encoded in color as a function of excitation polarization angle $\theta$ and energy $E$. **(b)** Differential reflectivity $\Delta R/R_{off}$ as a function of energy $E$ for the coupled mode $E_c$ and the uncoupled mode $E_{uc}$ in blue and green, respectively. Grey symbols represent the degree of linear polarization $\rho = \frac{\gamma_c - \gamma_{uc}}{\gamma_c + \gamma_{uc}}$ as a function of energy $E$. Red curves show corresponding simulations of both modes **(c)** Integrated differential reflectivity as a function of excitation polarization angle $\theta$ for the coupled mode $E_c$ and uncoupled mode $E_{uc}$ in blue and green, respectively.

Finally, we investigate the inter-particle coupling between the localised surface plasmon polaritons in the individual Au triangles, which form the bowtie nanoantenna [48] [30]. Therefore, we performed differential reflectivity measurements on an individual bowtie nanoantenna with $s_0 = 140nm$ and $g_0 = 10nm$ as a function of the excitation polarization angle $\theta$. Here, $\theta$ is defined as the angle between the electric field vector of the linearly polarized excitation and the long axis of the bowtie nanoantenna, i.e. the y-axis as defined in Figure 1 (a). We show in Figure 5 (a), the differential reflectivity signal $\Delta R/R_{off}$ of a single bowtie nanoantenna encoded in colour as a function of energy $E$ and excitation polarization angle $\theta$. We observe two energetically separated resonances at $E_c = 1.41 eV$ and $E_{uc} = 1.75 eV$ for $\theta_c = (0°, 180°)$ and $\theta_{uc} = (90°, 270°)$, respectively, which we attribute to the coupled and uncoupled nanotriangle resonances. The $E_c$-resonance appears at significantly lower energy as compared to the resonance of an individual nanotriangle $E_{uc}$ due to near-field coupling between the localised surface plasmon polaritons in the two triangles [48]. We support this assumption by numerical simulations of the scattering cross-section $\sigma$ for a bowtie nanoantenna with $s_0 = 140 nm$ and $g_0 = 10 nm$. We compare in Figure 5 (b) the simulation results for the coupled ($\theta_c = 0°$, red dashed curve) and uncoupled mode ($\theta_{uc} = 90°$, red solid curve) with the according $\Delta R/R_{off}$-spectra shown in blue and green, respectively. We obtain excellent qualitative and quantitative agreement between the experimental and theoretical resonance energies $(E_c^{ex}, E_c^{th}) = (1.41 eV, 1.40 eV)$ and $(E_{uc}^{ex}, E_{uc}^{th}) = (1.75 eV, 1.72 eV)$ and their according linewidth $(\Gamma_c^{ex}, \Gamma_c^{th}) = (0.36 eV, 0.36 eV)$ and $(\Gamma_{uc}^{ex}, \Gamma_{uc}^{th}) = (0.40 eV, 0.37 eV)$. Furthermore, we determine from the $\Delta R/R_{off}$-spectra of the coupled and uncoupled mode, the degree of linear polarization defined with respect to $I_c$ as $\rho = \frac{\gamma_c - \gamma_{uc}}{\gamma_c + \gamma_{uc}}$, where $I_c$ and $I_{uc}$ denote the corresponding differential reflectivity signals at $E_c$ and $E_{uc}$, respectively. In Figure 5 (b), we plot $\rho$ as a function of energy $E$ as grey symbols and observe for both the coupled and uncoupled mode clearly linearly polarized emission with $\rho_c = 80\%$ and $\rho_{uc} = -42\%$. Moreover, we plot in Figure 5 (c) the integrated $\Delta R/R_{off}$-signal at $E_c$ and $E_{uc}$ as a function of excitation polarization angle $\theta$ in blue and green, respectively. We observe a clear anti-correlation of the $\Delta R/R_{off}$-signal between the coupled and uncoupled mode, indicating that they scatter light along the long- ($\theta = 0°$) and short-axis ($\theta = 90°$) of the bowtie nanoantenna, respectively. From our findings, we conclude that the strong near-field coupling leads to a considerable red-shift of the localised surface plasmon resonance by $\Delta E \equiv |E_{uc} - E_c| = 340 meV$ as compared to uncoupled nanotriangles.

## Discussion

In summary, we presented a comprehensive study on the structural and optical properties of individual Au bowtie nanoantennas defined by electron beam lithographically on glass and GaAs substrates. The demonstrated highly uniform nanofabrication process in combination with the fast and reliable differential reflection spectroscopy established, paves the way for bowtie nanoantennas on high-refractive index semiconductor substrates [30] as essential building blocks in future optically active semiconductor-plasmonic integrated circuits [43] [50]. In particular the integration of antennas with other functional optical components such as for example plasmonic waveguides [51] [52] or photonic crystals [53] [52] requires along with this high degree of control and repeatability during nanofabrication also a fast, cheap and non-destructive spectroscopy method to independently test the optical response of the individual plasmonic units. Typical semiconductors like gallium arsenide and silicon rule out well-established techniques such as for example 'attenuated total internal reflection' [4] and 'transmission experiments' [2] since both require transparent substrates. Single-particle spectroscopy using a dark field microscope [41] requires immersion oils that contaminate the sample surface and, thus, modifies the optical properties of the plasmonic nanoparticles. In contrast to scanning near-field optical microscopy [40], which demands expensive equipment and records information in a serial manner, the demonstrated differential reflectivity spectroscopy offers quick and direct insights into the main optical properties of bowtie nanoantennas and potentially also works at cryogenic temperatures. The latter property becomes important when coupling plasmonic antennas to optically active emitters embedded in the semiconductor substrates [54] [55]. In combination with the control over the antenna position [56] and local electric contacts [13], this enables to engineer the spontaneous emission dynamics in such hybrid semiconductor-plasmonic nanosystems via the well-known Purcell-effect [57]. The enhancement is linked to the resonance linewidth $\Gamma_{res}$ via the so-called quality factor $Q = E_{res}/\Gamma_{res}$ and, therefore, nanostructures yielding minimum linewidth are favourable. Additional numerical simulations of truncated bowtie nanoantennas are presented in figure SM3, indicating a further improvement of $\Gamma_{res}$ by a factor $1.3 - 1.5 \times$. This is achieved by modifying the triangles of the bowtie nanoantenna to a 'two-wire gap'-like antenna [25] [13], giving rise to a reduction of the antenna volume, whilst simultaneously keeping the resonance energy constant. Further improvement of the $Q$-factor is expected by using single-crystalline metals due to a reduction of Ohmic losses in the metal as recently demonstrated in Ref. [58] [59]. Finally, it is well known that Ag instead of Au does not only allow to further increase the surface plasmon polariton energy, but also shows promise of decreased losses since the interband transitions are shifted towards higher energies [46]. In conclusion, we believe that our study provides an important step towards the marriage of semiconductor devices and nano-plasmonic concepts for the realization and optimization of efficient optical on-chip nanocircuits [60].

# References

[1] E. Dulkeith, A. C. Morteani, T. Niedereichholz, T. A. Klar, J. Feldmann, S. A. Levi, F. C. J. M. van Veggel, D. N. Reinhoudt, M. Möller and D. I. Gittins, "Fluorescence Quenching of Dye Molecules near Gold Nanoparticles: Radiative and Nonradiative Effects," *Phys. Rev. Lett.,* vol. 89, p. 203002, 2002.

[2] W. Rechberger, A. Hohenau, A. Leitner, J. R. Krenn, B. Lamprecht and F. R. Aussenegg, "Optical properties of two interacting gold nanoparticles," *Optics Communications,* vol. 220, p. 137, 2003.

[3] B. Lamprecht, G. Schider, R. T. Lechner, H. Ditlbacher, J. R. Krenn, A. Leitner and F. R. Aussenegg, "Metal Nanoparticle Gratings: Influence of Dipolar Particle Interaction on the Plasmon Resonance," *Phys. Rev. Lett.,* vol. 84, p. 4721, 2000.

[4] C. Sönnichsen, S. Geier, N. E. Hecker, G. von Plessen, J. Feldmann, H. Ditlbacher, B. Lamprecht, J. R. Krenn, F. R. Aussenegg, V. Z.-H. Chan, J. P. Spatz and M. Möller, "Spectroscopy of single metallic nanoparticles using total internal reflection microscopy," *Appl. Phys. Lett.,* vol. 77, p. 2949, 2000.

[5] K. B. Crozier, A. Sundaramurthy, G. S. Kino and C. F. Quate, "Optical antennas: Resonators for local field enhancement," *J. Appl. Phys.,* vol. 94, p. 4632, 2003.

[6] R. D. Grober, R. J. Schoelkopf and D. E. Prober, "Optical antenna: Towards a unity efficiency near-field optical probe," *Appl. Phys. Lett.,* vol. 70, p. 1354, 1997.

[7] D. A. Genov, A. K. Sarychev, V. M. Shalaev and A. Wei, "Resonant Field Enhancements from Metal Nanoparticle Arrays," *Nano Letters,* vol. 4, p. 153, 2004.

[8] E. Hao and G. C. Schatz, "Electromagnetic fields around silver nanoparticles and dimers," *J. Chem. Phys.,* vol. 120, p. 357, 2004.

[9] P. Biagioni, J.-R. Huang and B. Hecht, "Nanoantennas for visible and infrared radiation," *Rep. Prog. Phys.,* vol. 75, p. 024402, 2012.

[10] A. Sundaramurthy, K. B. Crozier, G. S. Kino, D. P. Fromm, P. J. Schuck and W. E. Moerner, "Field enhancement and gap-dependent resonance in a system of two opposing tip-to-tip nanotrianbles," *Phys. Rev. B,* vol. 72, p. 165409, 2005.

[11] A. G. Curto, G. Volpe, T. H. Taminiau, M. P. Kreuzer, R. Quidant and N. F. van Hulst, "Unidirectional Emission of a Quantum Dot Coupled to a Nanoantenna," *Science,* vol. 329, p. 930, 2010.

[12] D. P. Fromm, A. Sundaramurthy, P. J. Schuck, G. Kino and W. E. Moerner, "Gap-Dependent Optical Coupling of Single "Bowtie" Nanoantennas Resonant in the Visible," *Nano Letters,* vol. 4, p. 957, 2004.

[13] J. C. Prangsma, J. Kern, A. G. Knapp, S. Grossmann, M. Emmerling, M. Kamp and B. Hecht, "Electrically Connected Resonant Optical Antennas," *Nano Letters,* vol. 12, p. 3915, 2012.

[14] M. Kaniber, M. F. Huck, K. Müller, E. C. Clark, F. Troiani, M. Bicherl, H. J. Krenner and J. J. Finley, "Electrical control of the exciton-biexciton splitting in self-assembled InGaAs quantum dots," *Nanotechnology,* vol. 22, p. 325202, 2011.

[15] J. Kern, R. Kullock, J. Prangsma, M. Emmerling, M. Kamp und B. Hecht, „Electrically driven optical antennas," *Nature Photonics,* Bd. 9, p. 582, 2015.

[16] P. Biagioni, J. S. Huang, L. Duò, M. Finazzi and B. Hecht, "Cross Resonant Optical Antenna," *Phys. Rev. Lett.,* vol. 102, p. 256801, 2009.

[17] K. Kneipp, Y. Wang, H. Kneipp, L. T. Perelman, I. Itzkan, R. R. Dasari and M. S. Feld, "Single Molecule Detection Using Surface-Enhanced Raman Scattering (SERS)," *Phys. Rev. Lett.,* vol. 78, p. 1667, 1997.

[18] W. Srituravanich, L. Pan, Y. Wang, C. Sun, D. B. Bogy and X. Zhang, "Flying plasmonic lens in the near field for high-speed nanolithography," *Nature Nanotechnology,* vol. 3, p. 733, 2008.

[19] T. J. Silva, S. Schultz and D. Weller, "Scanning near-field optical microscope for the imaging of manetic domains in optically opaque materials," *Appl. Phys. Lett.,* vol. 65, p. 657, 1994.

[20] J.-M. Nam, C. S. Thaxton and C. A. Mirkin, "Nanoparticle-Based Bio-Bar Codes for the Ultrasensitive Detection of Proteins," *Science,* vol. 301, p. 1884, 2003.

[21] A. Tittl, X. Yin, H. Giessen, X.-D. Tian, Z.-Q. Tian, C. Kremers, D. N. Chirgin and N. Liu, "Plasmonic Smart Dust for Probing Local Chemical Reations," *Nano Letters,* vol. 13, p. 1816, 2013.

[22] M. Ringler, A. Schwemer, M. Wunderlich, A. Nichtl, K. Kürzinger, T. A. Klar and J. Feldmann, "Shaping Emission Spectra of Fluorescent Molecules with Single Plasmonic Nanoresonators," *Phys. Rev. Lett.,* vol. 100, p. 203002, 2008.

[23] A. Kinkhabwala, Z. Yu, S. Fan, Y. Avlasevich, K. Müllen and W. E. Moerner, "Large single-molecule fluorescence enhancements produced by a bowtie nanoantenna," *Nature Photonics,* vol. 3, p. 654, 2009.

[24] G. M. Akselrod, C. Argyropoulos, T. B. Hoang, C. Ciraci, C. Fang, J. Huang, D. R. Smith and M. H. Mikkelsen, "Probing the mechanisms of large Purcell enhancement in plasmonic nanoantennas," *Nature Photonics,* vol. 8, p. 835, 2014.

[25] P. Mühlschlegel, H.-J. Eisler, O. J. F. Martin, B. Hecht and D. W. Pohl, "Resonant Optical Antennas," *Science,* vol. 308, p. 1607, 2005.

[26] P. J. Schuck, D. P. Fromm, A. Sundaramurthy, G. S. Kino and W. E. Moerner, "Improving the Mismatch between Light and Nanoscale Objects with Gold Bowtie Nanoantennas," *Phys. Rev. Lett.,* vol. 94, p. 017402, 2005.

[27] H. A. Atwater and A. Polman, "Plasmonics for improved photovoltaic devices," *Nature Materials,* vol. 9, p. 205, 2010.

[28] J. M. Romo-Herrera, R. A. Alvarez-Puebla and L. M. Liz-Marzán, "Controlled assembly of plasmonic colloidal nanoparticle clusters," *Nansclae,* vol. 3, p. 1304, 2011.

[29] J. R. Krenn, G. Schider, W. Rechberger, B. Lamprecht, A. Leitner and F. R. Aussenegg, "Design of multipolar plasmon excitations in silver nanoparticles," *Appl. Phys. Lett.,* vol. 77, p. 3379, 2000.

[30] K. Schraml, M. Spiegl, M. Kammerlocher, G. Bracher, J. Bartl, T. Campbell, J. J. Finley and M. Kaniber, "Optical properties and interparticle coupling of plasmonic bowtie nanoantennas on a semiconducting substrate," *Phys. Rev. B,* vol. 90, p. 035435, 2014.

[31] H. Fischer and O. J. F. Martin, "Engineering the optical response of plasmonic nanoantennas," *Optics Express,* vol. 16, p. 9144, 2008.

[32] C. L. Haynes, A. D. McFarland, L. Zhao, R. P. Van Duyne, G. C. Schatz, L. Gunnarsson, J. Prikulis, B. Kasemo and M. Käll, "Nanoparticle Optics: The Importance of Radiative Coupling in Two-Dimensional Nanoparticle Arrays," *J. Phys. Chem. B,* vol. 107, p. 7337, 2003.

[33] N. Félidj, J. Grand, G. Laurent, J. Aubard, G. Lévi, A. Hohenau, N. Galler, F. R. Aussenegg and J. R. Krenn, "Multipolar surface plasmon peaks on gold nanotriangles," *J. Chem. Phys.,* vol. 128, p. 094702, 2008.

[34] S. R. K. A. A. Rodriguez, B. Maes, O. T. A. Janssen, G. Vechhi and J. Gómez Rivas, "Coupling bright and dark plasmonic lattice resonances," *Phys. Rev. X,* vol. 1, p. 021019, 2011.

[35] P. Lunnemann, I. Sersic and A. F. Koenderink, "Optical properties of two-dimensional magetoelectric point scattering lattices," *Phys. Rev. B,* vol. 88, p. 245109, 2013.

[36] C. M. Soukoulis, S. Linden and M. Wegener, "Negative Refractive Index at Optical Wavelengths," *Science,* vol. 315, p. 47, 2007.

[37] N. Yu, P. Genevet, M. A. Kats, F. Aieta, J.-P. Tetienne, F. Capasso and Z. Gaburro, "Light Propagation with Phase Discontinuities: Generalized Laws of Reflection and Refraction," *Science,* vol. 334, p. 333, 2011.

[38] N. Meinzer, W. L. Barnes and I. R. Hooper, "Plasmonic meta-atoms and metasurfaces," *Nature Photonics,* vol. 8, p. 889, 2014.

[39] S. A. Maier, Plasmonics: Fundamentals and Applications, New York: Springer, 2007.

[40] T. Klar, M. Perner, S. Grosse, G. von Plessen, W. Spirkl and J. Feldmann, "Surface-Plasmon Resonances in Single Metallic Nanoparticles," *Phys. Rev. Lett.,* vol. 80, p. 4249, 1998.

[41] C. Sönnichsen, T. Franzl, T. Wilk, G. von Plessen and J. Feldmann, "Drastic Reduction of Plasmon Damping in Gold Nanorods," *Phys. Rev. Lett.,* vol. 88, p. 077402, 2002.

[42] H. A. Atwater, "The Promise of Plasmonics," *Scientific American,* vol. 296, p. 56, 2007.

[43] R. Zia, J. A. Schuller, A. Chandran and M. L. Brongersma, "Plasmonics: the next chip-scale technology," *Materials Today,* vol. 9, p. 20, 2006.

[44] "FDTD Solutions," Lumerical Solutions Inc.;, [Online]. Available: https://www.lumerical.com/.

[45] J. Blakemore, "Semiconducting and other major properties of gallium arsenide," *J. Appl. Phys.,* vol. 53, p. R123, 1982.

[46] P. Billaud, J.-R. Huntzinger, E. Cottancin, J. Lermé, M. Pellarin, L. Arnaud, M. Broyer, N. Del Fatti and F. Vallée, "Optical extinction spectroscopy of single silver nanoparticles," *Eur. Phys. J. D,* vol. 43, p. 271, 2007.

[47] M. Maier and A. Wokaun, "Enhanced fields on large metal particles: dynamic depolarization," *Optics Letters,* vol. 8, p. 581, 1983.

[48] P. Nordlander, C. Oubre, E. Prodan, K. Li and M. I. Stockman, "Plasmon Hybridization in Nanoparticle Dimers," *Nano Letters,* vol. 4, p. 899, 2004.

[49] A. Wokaun, J. P. Gordon and P. F. Liao, "Radiation damping in surface-enhanced Raman scattering," *Phys. Rev. Lett.,* vol. 48, p. 957, 1982.

[50] V. J. Sorger, R. F. Oulton, R.-M. Ma and X. Zhang, "Toward integrated plasmonic circuits," *MRS Bulletin,* vol. 37, p. 728, 2012.

[51] G. Bracher, K. Schraml, M. Ossiander, S. Frédérick, J. J. Finley and M. Kaniber, "Optical study of lithographically defined, subwavelength plasmonic wires and their coupling to embedded quantum emitters," *Nanotechnology,* vol. 25, p. 075203, 2014.

[52] Z. Fang, L. Fan, C. Lin, D. Zhang, A. J. Meixner and X. Zhu, "Plasmonic Coupling of Bow Tie Antennas with Ag Nanowire," *Nano Letter,* vol. 11, p. 1676, 2011.

[53] E. R. Brown, C. D. Parker and E. Yablonovitch, "Radiation properties of a planar antenna on a photonic-crystal substrate," *J. Opt. Soc. Am. B,* vol. 10, p. 404, 1993.

[54] M. Pfeiffer, K. Lindfors, C. Wolpert, P. Atkinson, M. Benyoucef, A. Rastelli, O. G. Schmidt, H. Giessen and M. Lippitz, "Enhancing the Optical Excitation Efficiency of a Single Self-Assembled Quantum Dot with a Plasmonic Nanoantenna," *Nano Letters,* vol. 10, p. 4555, 2010.

[55] G. Bracher, K. Schraml, M. Blauth, J. Wierzbowski, N. Coca López, M. Bicherl, K. Müller, J. J. Finley and M. Kaniber, "Imaging surface plasmon polaritons using proximal self-assembled InGaAs quantum dots," *J. Appl. Phys.,* vol. 116, p. 033101, 2014.

[56] M. Pfeiffer, K. Lindfors, H. Zhang, B. Fenk, F. Phillipp, P. Atkinson, A. Rastelli, O. G. Schmidt, H. Giessen and M. Lippitz, "Eleven Nanometer Alignment Precision of a Plasmonic Nanoantenna with a Self-assembled GaAs Quantum dot," *Nano Letters,* vol. 14, p. 197, 2014.

[57] E. M. Purcell, "Spontaneous emission probabilities at radio frequencies," *Phys. Rev.,* pp. 69, 681, 1946.

[58] J. H. Park, P. Ambwani, M. Manno, N. C. Lindquist, P. Nagpal, S.-H. Oh, C. Leighton and D. J. Norris, "Single-Crystalline Silver Films for Plasmonics," *Adv. Mater.,* vol. 24, p. 3988, 2012.

[59] J.-S. Huang, V. Callegari, P. Geisler, C. Brüning, J. Kern, J. C. Prangsma, X. Wu, T. Feichtner, J. Ziegler, P. Weinmann, M. Kamp, A. Forchel, P. Biagioni, U. Sennhauser and B. Hecht, "Atomically flat single-crystalline gold nanostructures for plasmonic nanocircuitry," *Nature Communications,* vol. 1, p. 150, 2010.

[60] K. C. Y. Huang, M.-K. Seo, T. Sarmiento, Y. Huo, J. S. Harris and M. L. Brongersma, "Electrically driven subwavelength optical nanocircuits," *Nature Photonics,* vol. 8, p. 244, 2014.

[61] "Lumerical Solutions Inc.," [Online]. Available: https://kb.lumerical.com/en/particle_scattering_mie_3d.html.

[62] "Lumerical Solutions Inc.," [Online]. Available: https://kb.lumerical.com/en/ref_sim_obj_creating_rounded_corners.html.

[63] N. Liu, L. Langguth, T. Weiss, J. Kästel, M. Fleischhauer, T. Pfau and H. Giessen, "Plasmonic analogue of electromagnetically induced transparency at the Drude damping limit," *Nature Materials,* vol. 8, p. 758, 2009.

[64] T. Zhang, S. Callard, C. Jamois, C. Chevalier, D. Feng and A. Belarouci, "Plasmonic-photonic crystal coupled nanolaser," *Nanotechnology,* vol. 25, p. 345201, 2014.

## Methods

Sample fabrication and layout

The samples investigated were defined on semi-insulating GaAs [100] wafers or glass (MENZEL microscope cover slips) substrates. After cleavage, the samples were flushed with acetone and isopropanol (IPA). In order to get a better adhesion of the e-beam resist, the samples were put on a hot plate (170 °C) for 5 min. An e-beam resist (Polymethylmethacrylat 950 K, AR-P 679.02, ALLRESIST) was coated at 4000 rpm for 40s at an acceleration of 2000 rpm/s and baked out at 170°C for 300s, producing a resist thickness of $70 \pm 5$ nm. For the glass samples, we evaporated 10nm aluminium on top of the Polymethylmethacrylat layer to avoid charging effects during the e-beam writing. The samples were illuminated in a Raith E-line system using an acceleration voltage of 30kV and an aperture of 10μm. A dose test was performed for every fabrication run, as this crucial parameter depends on the varying e-beam current. Typical values were 800 μC/cm$^2$ for GaAs and 700 μC/cm$^2$ for glass substrates. After the e-beam writing the aluminium layer on the glass samples was etched away using a metal-ion-free photoresist developer (AZ 726 MIF, MicroChemicals). All samples were developed in Methylisobutylketon diluted with IPA (1:3) for 45s. To stop the development, the sample was rinsed with pure IPA. For the metalisation an e-beam evaporator was used to deposit a 5nm thick titanium adhesion layer for the glass and 35nm of gold for all substrates at a low rate of 1 Å/s. The lift-off was performed in 50°C warm acetone, leaving behind high-quality nanostructures with feature sizes on the order of 10nm.

Structural characterisation

To determine the geometrical parameters of the fabricated nanoantennas we took scanning electron microscopy images using a Raith E-line system at an electron acceleration voltage of $5kV$ and an aperture size of 10μm. We recorded the pictures by stepping from one antenna to another and conducting a single shot scan in order to avoid charging effects, which occur especially on the glass samples. The obtained images were analysed by hand using the "Carl Zeiss SmartTiff Annotation Editor" (V1.0.1.2). As stated in the main text we extracted $s$ and $g$ from high resolution scanning electron microscopy measurements of $\sim 300$ bowtie nanoantennas without any pre-selection. To quantify the tip radius we evaluated 20 "feed-gap tips" of the upper triangle and found a value of $r_c = 14 \pm 5$ nm.

Optical spectroscopy

For optical spectroscopy we used either a white-light super-continuum source (Fianium WhiteLase micro) for single particle studies or we collected and collimated the light from a halogen lamp (Philips Fibre Optic Lamp, Type 6423 XHP FO) for ensemble measurements. Both beams were sent through a beamsplitter and an apochromatic high numerical aperture (NA = 0.9) objective to focus the light onto the sample surface. We determined the spot sizes to be ($\emptyset_{spot}^{WhiteLase} = 1 \pm 0.1$μm) and ($\emptyset_{spot}^{halogen} \sim 5.2$μm), respectively. The sample were placed on an open-loop piezo stage (Thorlabs NanoMax) in combination with a tiltable stage in order to provide an accurate positioning and an exact alignment of the plane perpendicular to the optical path. The reflected light was collected by the same objective, transmitted through the beamsplitter, a fibre coupler and a multimode optical fibre before it was dispersed and analysed in a $0.5\text{m}$ imaging spectrometer (Princeton Instruments Acton SP2500i, grating: 300l/mm). Both excitation and detection channels were equipped with linear polarizers (Thorlabs, LBVIS100-MP2) and $\lambda/2$-waveplates (Thorlabs, AHWP10M-980) mounted on computer controlled motorized sample stages (Thorlabs, PRM1/MZ8) to adjust and analyse the polarization. For the measurements on glass (GaAs) we used a $600\text{nm}$ ($800\text{nm}$) long pass filters and a Si-CCD - Princeton Instruments Spec-10 (InGaAs linear array - Princeton Instruments, OMA V). When using the super continuum source to investigate the nanoantennas on GaAs, we also installed a $1064\text{nm}$ notch filter in order to suppress the residual light from the seed laser, which potentially can damage the InGaAs detector. To cover the broad energy range discussed in the main part of this work, we always recorded four spectra of different centre energies, which were merged afterward. The integration time was always set to $1\text{s}$.

Simulations

We simulated the scattering cross sections of the bowtie nanoantenna using a commercially available finite difference time domain solver (Lumerical Solutions, Inc., FDTD solutions, version: 8.11.387). The design of the simulation cell is based on the Mie scattering tutorial that can be found on the Lumerical homepage [61]. Consequently, we used a three dimensional simulation cell that is terminated by perfectly matched layers. The bowtie was modelled using the extruded N-sided equilateral polygon with rounded corners that is also provided on the Lumerical homepage [62]. To excite the structures we used a total field scattered field (TFSF) source and FDTD scattered field monitor to compute the scattering cross-section. At the centre of the simulation cell, i.e. around the bowtie feed-gap region, we used a mesh size of $2nm$, whereas in the outer regions the value was set to $4nm$. The used simulation file is available in the supplementary material.

# Acknowledgements

We acknowledge financial support of the DFG via the SFB 631, Teilprojekt B3 and the German Excellence Initiative via NIM. The authors gratefully acknowledge the support of the TUM International Graduate School of Science and Engineering (IGSSE).

# Author contributions

## Additional information

Supplementary information accompanies this paper at http://www.nature.com/scientificreports

## Competing financial interests:

The authors declare no competing financial interests.